\documentclass[twocolumn]{revtex4}
\usepackage{graphicx}
\begin{document}
\title{Wiring cost in the organization of a biological network}
\author{Yong-Yeol Ahn}
\affiliation{Department of Physics, Korea Advanced Institute of
  Science and Technology, Daejeon 305-701, Korea}
\author{Beom Jun Kim}
\affiliation{Department of Molecular Science and Technology, Ajou
  University, Suwon 442-749, Korea}
\author{Hawoong Jeong}
\affiliation{Department of Physics, Korea Advanced Institute of
  Science and Technology, Daejeon 305-701, Korea}
\date{\today} 
\begin{abstract}

To find out the role of the wiring cost in the organization of the neural network of the nematode \textit{Caenorhapditis elegans} (\textit{C. elegans}), we build the neuronal map of \textit{C. elegans} based on geometrical positions of neurons and define the cost as inter-neuronal Euclidean distance \textit{d}. We show that the wiring probability decays exponentially as a function of \textit{d}. Using the edge exchanging method and the component placement optimization scheme, we show that positions of neurons are not randomly distributed but organized to reduce the total wiring cost. Furthermore, we numerically study the trade-off between the wiring cost and the performance of the Hopfield model on the neural network.

\end{abstract}
\maketitle
Human beings have been attracted by their own brains' vast ability and complexity~\cite{neuroscience}. Although the topology of connections between neurons in a brain is the basic interest in the brain research, it is impossible to obtain the whole connections of human brain directly because the number of neurons is so huge (more than hundreds of billions). Only macroscopic and indirect informations on connectivity are accesible in the entire brain's scale so far~\cite{Eguiluz03, Carmichael94, Felleman91, Lewis00}. However, primitive organisms that have rather small sizes shed a light on the research of the structure of connections in a brain thanks to their relatively simple organization.

A nematode \textit{Caenorhabditis elegans} (\textit{C. elegans}), which was the first multicellular organism whose whole genome was sequenced in 1998~\cite{CSC}, is the one of such examples. The remarkable feature of this nematode is that most individuals of \textit{C. elegans} have the same cellular properties such as shapes, connectivities and positions of neurons, which makes it possible to label each neuron. Throughout the pioneering work by White \textit{et al.}, the connectivity of the neuronal network has been uncapped completely~\cite{White86}. 

A decade after the intensive works on the nematode, the researches on large-scale complex networks were initiated. Since the founding works of modeling complex networks~\cite{Watts98,Barabasi99} and the revealing of the real networks' structures~\cite{Albert99}, numerous models have been suggested and studied, and various real networks have also begun to reveal their characteristics, such as small-worldness, power-law degree distribution, and high level of clustering~\cite{GeneralReview}. The concepts, tools, and methods of the complex network researches have been diffused into various fields, especially into biology.

As the knowledge about biological networks grows, the issue of identifying the organizing principle of them becomes more substantial. In this issue, a powerful approach is focusing on the high effectiveness of biosystems because of the very essentiality of natural selection in the evolution. There are evidences of optimization in biological networks as we see in the normal biological organs. For example, the average distance between two metabolites of 43 microorganisms' metabolic networks remain the same even though their numbers of nodes span from 200 to 800~\cite{Jeong00}. This optimization viewpoint can be also applied to technological systems (e.g. electric circuit~\cite{Cancho01}, modular structure of a software~\cite{Cancho01a}) because technological systems also evolve by the pressure toward high efficiency. It was discovered that the power-law degree distribution can come out from the optimization based models~\cite{Cancho01,Valverde02, YYAhnPre}.  In these models, a total cost is defined as the number of links (or total length of links) and the effectiveness is defined by several different measures, such as the short average distance or the large clustering coefficient. The energy function to be optimized is defined using the cost and the effectiveness measures, and then, by minimizing the energy function, the scale-free network emerges as a result. 

Evolution has built the current state of the brain. The main function of the brain is the capability to recognize and react flexibly to the various patterns of surrounding situations because it greatly helps survival and reproduction of an organism. In this context, the pattern recognition ability of Hopfield model~\cite{Hopfield} was adopted as a measure for the functionality of a neuronal network~\cite{BJKim04}. However, when the performance of a neuronal network is only defined by the pattern recognition ability, the neuronal network of \textit{C. elegans} shows worse performance than Barabasi-Albert (BA) network~\cite{Barabasi99} and Watts-Strogatz (WS) network~\cite{Watts98} with the rewiring probability $p=1.0 $~\cite{BJKim04}. Because they are random network models which are not designed for the neural computation, this result may imply two different conclusions: The measure is invalid, or the measure reflects the true performance of the neural network but there are other constraints which obstruct the pressure to make a high performance neural network. 

A probable constraint is the cost. We need to build a very complex (expensive) structure for a well-functioning brain and it consumes a large amount of energy. Therefore, the efficiency of it highly limits its function~\cite{Laughlin03}.  Furthermore, the strong evidence of the cost optimization in \textit{C. elegans} was discovered by showing that the natural position of ganglia in \textit{C. elegans} has the lowest wiring cost among the vast number of all possible positioning combinations~\cite{Cherniak94}. Here, we construct the three-dimensional neuronal map of \textit{C. elegans}, and we display and estimate the effect of the cost against the effect of the performance in the organization of the neural network of \textit{C. elegans}.

\begin{figure}
  \resizebox{80mm}{!}{\includegraphics{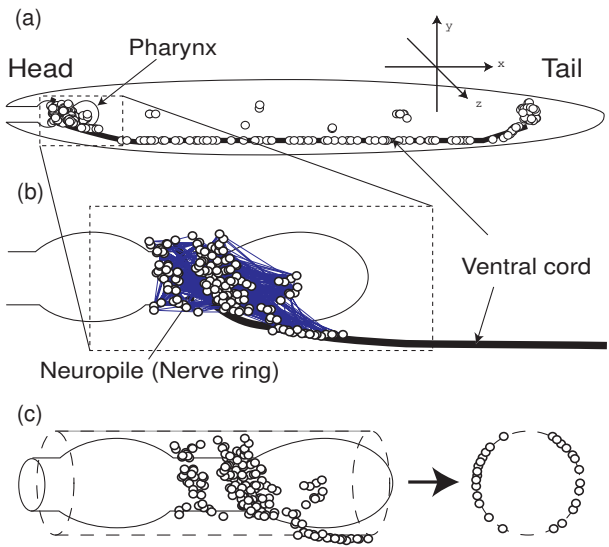}}
  \caption{ (a) The schematic diagram of the locations of the neurons in \textit{C. elegans} (b) The locations of head neurons, neuropile, and ventral cord and the connections between the head neurons. (c) The schemetaic diagram of constructing the three-dimensional neuronal map from a two-dimensional figure of the head neurons.}
  \label{map}
\end{figure}

First of all, we construct the map of neurons in \textit{C.  elegans}. We define the positions of neurons as that of the soma which is collected from the figures of the actual nervous system~\cite{White86}. We introduce following assumptions to construct three-dimensional map from the two-dimensional figures. The first assumption is that the head neurons wrap the pharynx closely and they are located on the surface of an imaginary cyllinder whose diameter is the same as the pharynx's diameter. At the center of the body, we put the neurons that appear in the both figures from the viewpoint of left-side and right-side. We place the ventral cord neurons at the bottom center of the body. For the tail neurons, we assume that they are on the line $y=z$ and $y=-z$. We assume that the remaining body neurons are placed just below the cuticle layer due to the pseudocoelome (body cavity) of the organism. The positions of neurons and the schematic diagram of map construction is shown in Fig. \ref{map}.

Although the wiring is generally not straight and guided by the structures such as the nerve ring and the ventral cord, we start here with the null hypothesis that approximates the wiring cost between any two neurons is only proportional to the Euclidean distance between them and the total cost is defined by the sum of the costs of all connected neuron pairs (the head-tail length of the worm is normalized to 1 for convenience).

\begin{figure}
  \resizebox{80mm}{!}{\includegraphics{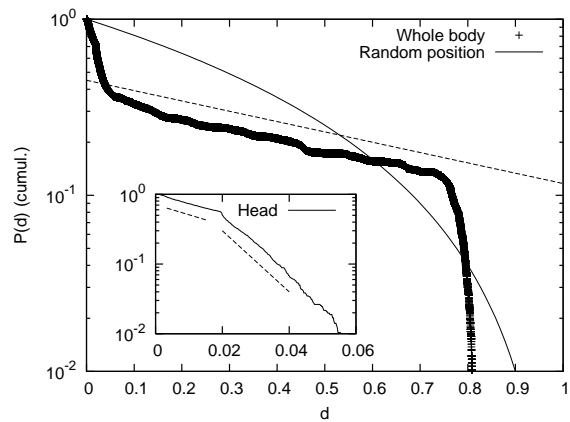}}
  \caption{The cumulative plot of inter-neuron distance (d) distribution in     \textit{C. elegans} (The non-cumulative distribution is calculated by differentiating the cumulative distribution). The distribution shows piecewise exponential decay. We plot the case that all neurons are randomly distributed through the body (``Random position''). This distribution has the functional form of $(d-1)^2$. If we ignore the extremely long-range links ($d>0.75$), the original network has less long links and much larger amount of short links in comparison with this random distribution. The inset shows the distance distribution of the neuron pairs only in the head. The kink is made by the gap between left part of the head neurons and the right part of them, which corresponds to the radius of pharynx, 0.02. If we make a projection of all the neurons onto the plane that is orthogonal to $x$ axis, the kink vanishes. }
  \label{whole}
\end{figure}

The cost distribution of the original \textit{C. elegans} network is shown in Fig. \ref{whole}. Through the whole body, the cost distribution exhibits piecewise exponential decaying distribution with three different length scales: 0.03, 0.01, and 0.74. \textit{i.e.}, each piece decays as $ e^{-d/\xi} $ where $\xi$ is 0.03, 0.01 and 0.74 for each. The longest length scale 0.74 is same as the length scale where body-spanning long link appears. We believe that the other two length scales in the head are related to the radius of pharynx ($=0.02$). We draw the distribution made by randomly distributing the positions of all the neurons anywhere in the body and compare it with the original neural network's distribution. Note that, in the original neural network, there are certain amount (about 10\%) of long range links which span about 80\% of the whole body and these body-spanning wirings must be crucial for the coordinated function of the worm, such as coherent movement and the egg-laying behavior, as is the spinal cord in our body. Besides these long range links, the neural network has smaller number of middle range links and much larger number of short links, which supports our proposition that the wiring cost is an important factor in \textit{C. elegans}. This result also implies the existence of the relation between the connection probability and the Euclidean distance of neuron pairs although the connections are not straight in reality, as suggested in~\cite{Kaiser04}. 

\begin{figure}
  \resizebox{80mm}{!}{\includegraphics{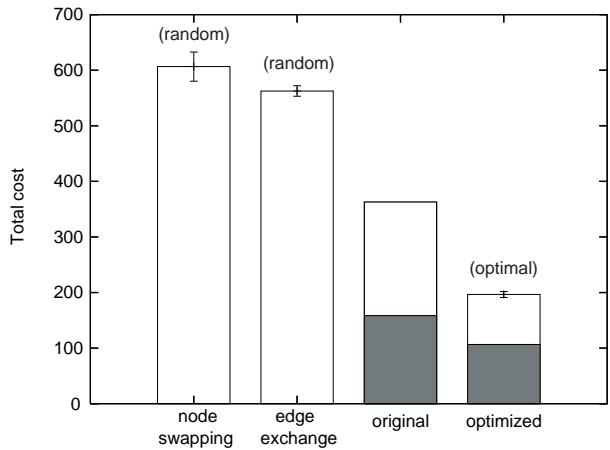}}
  \caption{\textit{C. elegans} neuronal network's cost is compared with
    randomly shuffled networks and an optimized network. `Node swapping' stands
    for the cost of the network which the nodes are randomly swapped, 
    `edge exchange' stands for that of which the edges are randomly exchanged,
    and the `optimized' stands for the network whose cost is optimized by node-swapping method. The dark box shows the cost of the network whose long range links ($>0.75$) are removed. The standard deviation is shown.}
  \label{random}
\end{figure}

To seek further evidences for the role of the wiring cost, we introduce two variational methods. The connections (neural circuit's topology) and the positions are two of the most important entities in a neural network and hence two methods vary either of them. The first method is the edge exchange method suggested in~\cite{Maslov02}, which conserves the positions and degrees of nodes, but changes individual links randomly. The second method is the node swapping method that is used in~\cite{Cherniak94} and named as \textit{Component Placement Optimization} (CPO), which randomly chooses two nodes and swap the positions while preserving all the connections (Note that the word `optimization' may mislead. We use this method not only for optimizing a network but also randomizing a network and we call it ``node swapping method'' henceforth). We randomly shuffle the original network using both methods and compare the resulting networks with the original one. In addition, we also compare them with the network that is optimized with Monte Carlo method. We start with high temperature and slowly drop the temperature. We draw the result in the Fig \ref{random}. The original network's total cost is smaller than that of the randomly shuffled networks, which supports that there exists the role of the cost in the organization of the \textit{C. elegans} neural network. Moreover, if we remove body-spanning links ($> 0.75$ ), the cost ratio of the original network to the optimized one is enhanced from 0.53 to 0.67. It means there are lots of long range links which have to be placed without consideration of the cost. 

\begin{figure}
  \resizebox{80mm}{!}{\includegraphics{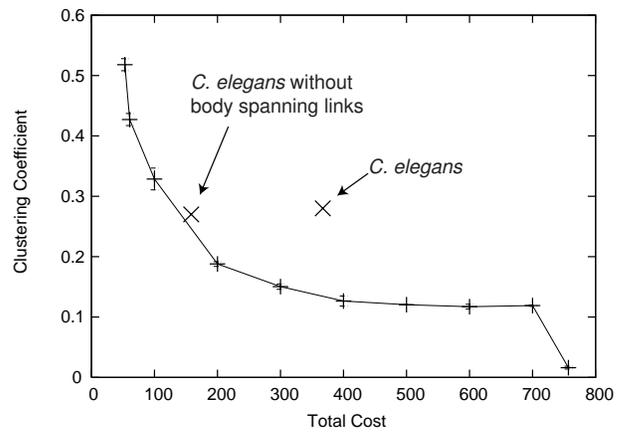}}
  \caption{The total cost versus the clustering coefficient. We make networks which have a designated total cost and calculate the clustering coefficient of them.  }
  \label{tradeoff2}
\end{figure}

To observe the relation between the performance and the cost more systematically, we make a randomly shuffled network which have a desired total wiring cost. The result is shown in Fig. \ref{tradeoff2}. This figure shows the negative correlation between the total cost and the clustering coefficient. We observe that the network whose cost is minimized by the edge exchange method has the clustering coefficient of 0.43, which is much larger than the original network (0.28). And we also observe that the network, whose cost is maximized by the same method, shows pretty low clustering coefficient (0.02) than the original. Note that the \textit{C. elegans}'s neuronal network without body-spanning long links has similar clustering coefficient with the cost-tuned random network. It means that the clustering coeffient is not an important factor in the organization of \textit{C. elegans} neuronal network and those body spanning links don't contribute to the clustering coefficient.

We show that although the \textit{C. elegans} neural network is not a cost-optimal network, the network is far from random. The connection probability follows exponential decaying law of Euclidean distance, and if we neglect body-spanning long links, which may be considered as a spinal cord in human, the total cost approaches to the position-optimal network and the clustering coefficient approaches to that of the cost-tuned random network. We cannot find the evidence that the clustering coefficient affects the structure of the neuronal network. But, we find that the body-spanning links (20\% of all links) do not increase the clustering coefficient much. Because the high level of clustering increases the path length in general (WS network and power grid), this fact may impliy that those body-spanning links are designed to connect all neurons throughout the body effectively. And although the effect of clustering coeffient in the organization of the neuronal network cannot be found, we show that the total wiring cost competes with the clustering coefficient that is closely related to the pattern retrieval performance of Hopfield model. 

We want to suggest three reasons why the \textit{C. elegans} neural network is not cost-optimal. First possible reason is the existence of the neuropile, such as the nerve ring and ventral cord, which is expected to reduce the wiring cost between distant neurons. Second one is the functional constraints. The minimal subnetwork topologies for generating certain dynamical behaviors such as oscillation or chaotic motion are identified and named as  ``dynamical motifs''~\cite{Zhigulin03}. This result shows that some network topologies are unavoidable to perform certain behaviors and these essential topologies may obstruct the wiring cost optimization. The last reason is the developmental constraints. The developmental plan of every living organism is highly conserved through the history of life~\cite{Wolpert02}, which means the plan can be changed hardly and hence is a highly optimized process. Because more complex development process needs more cost and has higher risk of failure, an organism should makes the development process as simple as possible. That is, even if \textit{C. elegans} can minimize the wiring cost by modifying the developmental process, the minimization may not happen because of the developmental cost. Although the cost is also an important issue in the human brain, our developmental program does not control the development of every single neuron~\cite{Kalisman05}.

We conclude that although the wirings generally follow non-optimal paths, the wiring probability is largely determined by the Euclidean distance between the neurons and the wiring cost is an important factor that affects the development of neural network of \textit{C. elegans} and competes with the clustering coeffient.

\acknowledgments{
Y.-Y. A. was supported by the Ministry of Science and Technology through Korean Systems Biology Research Grant (M10309020000-03B5002-00000), and H. J. was supported by KOSEF ABRL Program through Grant No. R14-2002-059-01002-0. B. J. K. was supported by the Ministry of Science and Technology through the Nanoscopia Center of Excellence at Ajou University.}


\begin{references}
\bibitem{neuroscience} M. Bear, M. F. Bear, B. W. Connors, and M. A. Paradiso, \textit{Neuroscience: Exploring the Brain} (Lippincott Williams \& Wilkins, 2002)

\bibitem{Eguiluz03} V. M. Egu\'iluz, D. R. Chialvo, G. A. Cecchi, M. Baliki, and A. V. Apkarian, Phys. Rev. Lett. \textbf{94}, 018102 (2005).

\bibitem{Carmichael94} S. T. Carmichael, and J. L. Price, J. Comput. Neurol. \textbf{346}, 366 (1994).

\bibitem{Felleman91} D. J. Felleman, and D. C. van Essen, Cereb. Cortex \textbf{1}, 1 (1991).

\bibitem{Lewis00} J. W. Lewis, and D. C. van Essen, J. Comput. Neurol. \textbf{428}, 79 (2000).

\bibitem{CSC} The \textit{C. elegans} Sequencing Consortium, Science
  \textbf{282}, 2012 (1998).

\bibitem{White86} J. G. White, E. Southgate, J. N. Thomson, and
  S. Brenner, Phil. Trans. R. Soc. Lond. Ser. B \textbf{314}, 1 (1986).

\bibitem{Watts98} D. J. Watts and S. H. Strogatz, Nature (London)
  \textbf{393}, 440 (1998).

\bibitem{Barabasi99} A.-L. Barab\'asi and R. Albert, Science
  \textbf{286}, 509 (1999);
  A.-L. Barab\'asi, R. Albert, and H. Jeong,
  Physica A \textbf{272}, 173 (1999).

\bibitem{Albert99} R. Albert, H. Jeong,
  and A.-L. Barab\'asi Nature \textbf{401}, 130-131 (1999).

\bibitem{GeneralReview} R. Albert and A.-L. Barab\'asi, 
  Rev. Mod. Phys. \textbf{74}, 47 (2002);
  S. N. Dorogovtsev and J. F. F. Mendes,
  Adv. Phys. {\bf 51}, 1079 (2002); 
  M. E. J. Newman, 
  SIAM Rev. \textbf{45}, 167 (2003).

\bibitem{Jeong00} H. Jeong, B. Tomber, R. Albert, Z. N. Oltvai, and
  A.-L. Barabasi, Nature (London) \textbf{407}, 651 (2000).

\bibitem{Cancho01} R. F. Cancho, C. Janssen, and R. V. Sol\'e,
 Phys. Rev. E \textbf{64}, 046119 (2001).

\bibitem{Cancho01a} R. F. Cancho and R. V. Sol\'e,
  e-print cond-mat/0111222.

\bibitem{Valverde02} S. Valverde,  R. F. i Cancho, and R. V. Sol\'e,
  Europhys. Lett. \textbf{60}(4), 512 (2002).

\bibitem{YYAhnPre} Y.-Y Ahn, P.-J. Kim, and H. Jeong, to be published.

\bibitem{Hopfield} See for reviews, e.g., J. Hertz, A. Krogh, and R. G. Palmer, \textit{Introduction to the Theory of Neural Computation} (Perseus Books, Cambridge, 1991). 

\bibitem{BJKim04} B. J. Kim,
  Phys. Rev. E \textbf{69}, 045101(R) (2004).

\bibitem{Laughlin03} S. B. Laughlin and T. J. Sejnowski,
  Science \textbf{301}, 1870 (2003).

\bibitem{Cherniak94} C. Cherniak, J. Neurosci. \textbf{14},
  2408 (1994).

\bibitem{Kaiser04} M. Kaiser and C. C. Hilgetag, Neurocomputing \textbf{58-60}, 297 (2004)

\bibitem{Maslov02} S. Maslov and K. Sneppen, Science \textbf{296}, 910 (2002)

\bibitem{Zhigulin03} V. P. Zhigulin, Phys. Rev. Lett. \textbf{92}, 238701 (2004).

\bibitem{Wolpert02} L. Wolpert, R. Beddington, T. Jessell, P. Lawrence, E. Meyerowitz, and J. Smith, \textit{Principles of Development} (Oxford University Press, New York, 2002). 

\bibitem{Kalisman05} N. Kalisman, G. Silberberg and H. Markram, Proc. Natl. Acad. Sci. U.S.A. \textbf{102}, 880 (2005).


\end{references}
\end{document}